# Metabolomic profiles in Jamaican children with and without autism spectrum disorder

Akram Yazdani[1,2], Maureen Samms-Vaughan[3], Sepideh Saroukhani[1,2], Jan Bressler[4], Manouchehr Hessabi[2], Amirali Tahanan[2], Megan L. Grove[4], Tanja Gangnus[5], Vasanta Putluri[6], Abu Hena Mostafa Kamal[5,6], Nagireddy Putluri[5], Katherine A. Loveland[7], Mohammad H. Rahbar[1,2,8*]

1. Division of Clinical and Translational Sciences, Department of Internal Medicine, McGovern Medical School, The University of Texas Health Science Center at Houston, Houston, Texas, USA
2. Biostatistics/Epidemiology/Research Design (BERD) Component, Center for Clinical and Translational Sciences (CCTS), The University of Texas Health Science Center at Houston, Houston, Texas, USA
3. Department of Child & Adolescent Health, The University of the West Indies (UWI), Mona Campus, Kingston 7, Jamaica
4. Human Genetics Center, Department of Epidemiology, School of Public Health, The University of Texas Health Science Center at Houston, Houston, Texas, USA
5. Department of Molecular and Cellular Biology, Baylor College of Medicine, Houston, Texas, United States.
6. Advanced Technology Core, Dan L. Duncan Comprehensive Cancer Center, Baylor College of Medicine, Houston, Texas, United States.
7. Louis A Faillace, MD, Department of Psychiatry and Behavioral Sciences, McGovern Medical School, The University of Texas Health Science Center at Houston, Houston, Texas, USA
8. Department of Epidemiology, School of Public Health, The University of Texas Health Science Center at Houston, Houston, Texas, USA

*Correspondence:
Mohammad H. Rahbar

**Background:** Autism spectrum disorder (ASD) is a complex neurodevelopmental condition with a wide range of behavioral and cognitive impairments. While genetic and environmental factors are known to contribute to its etiology, the underlying metabolic perturbations associated with ASD which can potentially connect genetic and environmental factors, remain poorly understood. Therefore, we conducted a metabolomic case-control study and performed a comprehensive analysis to identify significant alterations in metabolite profiles between children with ASD and typically developing (TD) controls.

**Objective:** To elucidate potential metabolomic signatures associated with ASD in children and identify specific metabolites that may serve as biomarkers for the disorder.

**Methods:** We conducted metabolomic profiling on plasma samples from participants in the second phase of Epidemiological Research on Autism in Jamaica (ERAJ-2), which was a 1:1 age (±6 months)-and sex-matched cohort of 200 children with ASD and 200 TD controls (2-8 years old). Using high-throughput liquid chromatography-mass spectrometry techniques, we performed a targeted metabolite analysis, encompassing amino acids, lipids, carbohydrates, and other key metabolic compounds. After quality control and imputation of missing values, we performed univariable and multivariable analysis using normalized metabolites while adjusting for covariates, age, sex, socioeconomic status, and child's parish of birth.

**Results:** Our findings revealed unique metabolic patterns in children with ASD for four metabolites compared to TD controls. Notably, three of these metabolites were fatty acids, including myristoleic acid, eicosatetraenoic acid, and octadecenoic acid. Additionally, the amino acid sarcosine exhibited a significant association with ASD.

**Conclusions:** These findings highlight the role of metabolites in the etiology of ASD and suggest opportunities for the development of targeted interventions.

## Introduction

Autism spectrum disorder (ASD) is a complex neurodevelopmental condition with higher than 1% worldwide prevalence, and there has been an observable upward trend in this rate over the last decade[1]. The diagnosis of ASD is conducted by behavior assessment due to the limited knowledge of the biological mechanisms governing its etiology. Therefore, understanding the molecular underpinnings of ASD holds significant promise for the identification of novel diagnostic and treatment strategies[2].

Metabolomics is one emerging field of study that has become a valuable tool in understanding the intricate biochemical signatures associated with ASD. Detecting the metabolic alterations linked to ASD has the potential to enhance the accuracy of early diagnosis, but also offers valuable insights into the disrupted underlying biological pathways in ASD[3]. We conducted metabolomic profiling on plasma samples from participants in the second phase of Epidemiological Research on Autism in Jamaica (ERAJ-2) study, which was a 1:1 age (±6 months)-and sex-matched cohort of 200 children with ASD and 200 TD controls (2-8 years old). Using high-throughput liquid chromatography-mass spectrometry techniques, we performed a targeted metabolite analysis, encompassing amino acids, lipids, carbohydrates, and other key metabolic compounds. We first conducted quality control and imputation of missing values, a step we previously discussed regarding its importance[4]. We then investigated the alteration of metabolites associated with ASD. We performed the analysis using both univariable and multivariable analyses. The latter is a regularized model for simultaneous analysis of 96 metabolites, which was optimized using cross-validation techniques while considering the accuracy. All analyses in this study were adjusted for age, sex, socioeconomic status, and child's parish of birth.

This study aims to uncover potential metabolic signatures associated with ASD in Jamaican children, offering insights into its underlying biological mechanisms that may contribute to early diagnosis of ASD.

## Materials and Methods

### Subjects

This study comprises data from 200 pairs of case-control individuals enrolled in phase 1 and 2 of The Epidemiological Research on Autism in Jamaica (ERAJ), a focused case-control study targeting children aged 2 to 8 years. Initiated in December 2009, the study invited children at risk for ASD identified based on Diagnostic and Statistical Manual of Mental Disorders (DSM-IV-TR) criteria[5] and the Childhood Autism Rating Scale[6]. Confirmatory evaluations for ASD cases involved standardized tools, including the Autism Diagnostic Observation Schedule (ADOS)[7], ADOS-2[8], and the Autism Diagnostic Interview-Revised (ADI-R)[9]. For each confirmed ASD case, a typically developing (TD) control was recruited from schools or well-child clinics whose age was within six months of age. TD status was verified using the Social Communication Questionnaire (SCQ)[10], adhering to guidelines (SCQ score of 0–6). After completing the questionnaires, blood samples were collected from participants without requiring fasting. All parents provided written informed consent, and when applicable, an assent was collected from 7-8-year-old children before they participated in this ERAJ study. Detailed recruitment and assessment procedures for both ASD cases and controls can be found in prior references[11].

All procedures performed in studies involving human participants were in accordance with the ethical standards of the institutional and/or national research committee and with the 1964 Helsinki Declaration and its later amendments or comparable ethical standards. The ERAJ study protocol has been approved by the Institutional Review Boards (IRBs) of both the University of Texas Health Science Center at Houston (UTHealth) (IRB Protocol number: HSC-SPH-09-0059) and The University of the West Indies (UWI), in Jamaica.

**Plasma samples collection and metabolite extraction**

For targeted metabolomics analysis, the human serum samples were thawed, including mouse liver pool as Quality Control (QC), and were mixed with 750 µL of internal standard (ISTD) mix in methanol-water (4:1). Metabolites were extracted using the liquid-liquid extraction method described previously[12,13]. Following partitioning with ice-cold chloroform and water, the organic and aqueous layers were meticulously transferred into new glass vials. Proteins and lipids were removed from extracted samples using a 3K Amicon-Ultra filter (Millipore Corporation, Billerica, MA). The extracted total metabolites samples were analyzed through high-throughput Liquid Chromatography-Mass Spectrometry (LC-MS/MS) techniques described previously[13–15].

The chromatographic separation of extracted metabolites was performed through Hydrophilic Interaction Chromatography (HILIC) techniques. The metabolites were separated through the XBridge Amide HPLC column (3.5 µm, 4.6 x 100 mm, Waters, Milford, MA) in both electrospray ionization (ESI) positive and negative mode. In positive ionization mode, the flow rate was set to 0.3 mL/min and an injection volume of 5 µL was applied. In ESI negative mode, the analysis employed a solvent flow rate of 0.3 mL/min with an injection volume of 10 µL.

For the analysis of fatty acids, a Luna 3 µm Phenyl-Hexyl column (150 × 2 mm; Phenomenex, Torrance, CA) was utilized. Mobile phases A and B consisted of 10 mM ammonium acetate (pH 8) and methanol. The gradient flow was as follows: 0-8 min 40% B, 8-13 min 50% B, 13-23 min 67%, 23-30 min 100%, and 30 min 40%, followed by re-equilibration until the end of the 37 min gradient to the initial starting condition of 40% B. The flow rate of the solvents used for analysis was 0.2 mL/min, with an injection volume of 20 µL.

The above-mentioned volume of samples was injected and the data was acquired via multiple reaction monitoring (MRM) using a 6495 Triple Quadrupole mass spectrometry coupled to an HPLC system (Agilent Technologies, Santa Clara, CA) through Agilent Mass Hunter Software.

**Data preprocessing**

The serum metabolites data were extracted from two sets, each comprising 100 pairs. Metabolomics extraction was independently performed for each set, with samples randomly assigned to 4 batches. The acquired data were analyzed and integration of each peak was performed using Agilent Mass Hunter Quantitative Analysis software. The relative peak area normalized to the internal standard was log2 transformed and batch correction was carried out using the ComBat Package[16].

To combine the two sets of data, we performed batch effect correction using the first principal component (**Figure S1A, B**). Quality assessment included clustering metabolites for both cases and controls, revealing comparable cluster formations in both groups (**Figure S2A-B**)

**Statistical methods**

We performed a multivariable analysis to simultaneously analyze 96 metabolites while adjusting for covariates. The model incorporates a regularization term with two parameters, mixing and regularization[17]. The optimal mixing parameter was determined based on the highest accuracy achieved by models when the samples were divided into training and test sets (70% and 30%, respectively). For the regularization parameters, optimization was achieved through 10-fold cross-validation, selecting the value that minimized the partial likelihood deviance from the model.

To ascertain the robustness of our findings, we estimated empirical 95% confidence intervals (CIs) for the metabolomics coefficients using bootstrapping over 500 iterations[18]. If the 95% CI for any coefficient did not encompass the origin, we considered the perturbation of the corresponding metabolites to be empirically significant.

We also conducted a univariate analysis using a generalized linear regression model, accounting for covariates (age, sex, socioeconomic status, and the child's parish of birth). The resulting *p*-values underwent correction for multiple comparisons through the false discovery rate (FDR) method[19]. Significance was attributed to perturbations between cases and controls with an FDR < 0.05.

Given that our samples were not collected under fasting conditions, paired sample analysis was not performed, as it could introduce bias and confound the interpretation of results. In addition, one of the samples was excluded from the analysis due to the absence of metabolomics data.

**Results**

We initially investigated the dietary habits of both cases and controls and assessed potential distinctions in their food consumption patterns. As described in detail in **Table S1**, this analysis revealed significant distinctions in certain categories of food consumption between the two groups. Therefore, we further explored the association between metabolites and food consumption through correlation analysis (**Figure 1A**). Since our study found either no or weak correlations between metabolites and food consumption ($r^2 < 0.06$), we made the decision not to include any scores for food consumption in the subsequent analysis.

We then applied a regularized multivariable model to simultaneously investigate perturbations of all metabolites, comparing the children with ASD to TD children. This model identified 44 metabolites with non-zero coefficients while adjusting for covariates (age, sex, socioeconomic status, and the child's parish of birth). Among these metabolites, 4 were significant based on the estimated CI for the adjusted odds ratio (AOR) (**Figure 1B**); myristoleic acid (AOR 1.96, 95% CI 1.62-3.49), eicosatetraenoic acid (AOR 0.52, 95% CI 0.96-0.72), octadecenoic acid (ODCA) (AOR 1.57, 95% CI 1.07-3.78), and sarcosine/alanine (AOR 1.79, 95% CI 1.34-3.35).

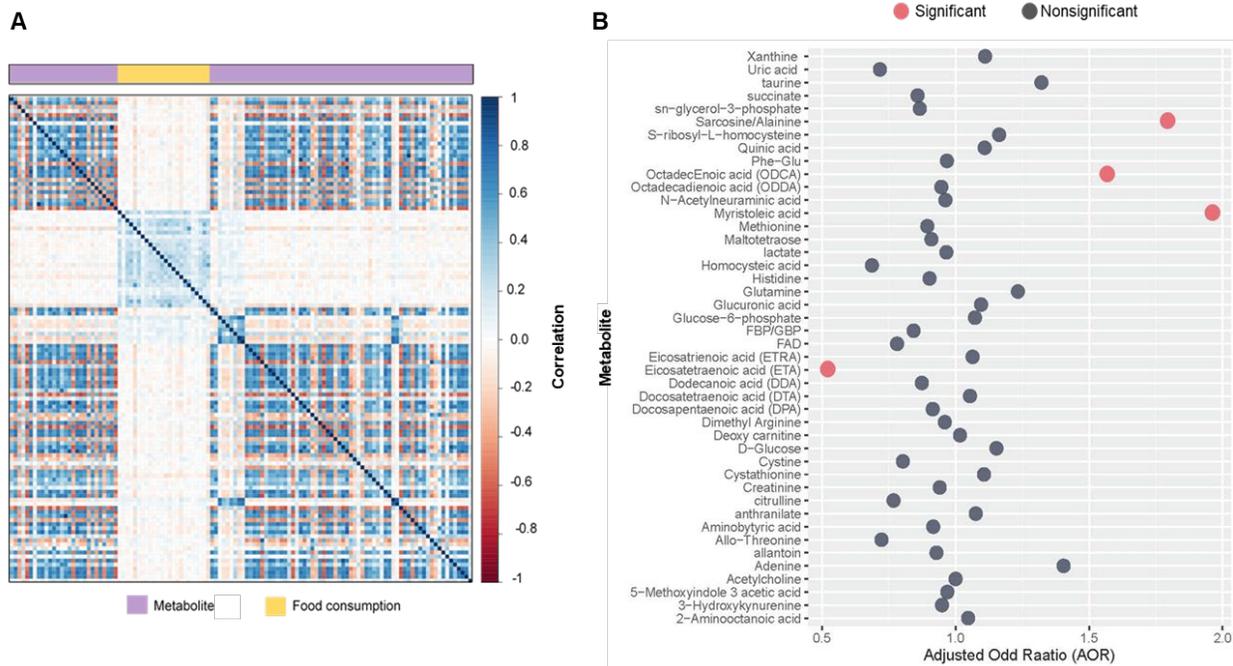

**Figure 1. A.** Representing either no or weak correlations between metabolites and food consumption. **B.** Adjusted odds ratio of metabolites with nonzero coefficients based on a regularized multivariable model. The red dot represents the adjusted odds ratio of metabolites empirically significant using the estimated confidence interval.

In addition, we conducted a univariable analysis while adjusting for covariates. Following correction for multiple testing (FDR < 0.05), myristoleic acid was significantly elevated among the children with ASD (AOR 1.41, 95% CI 1.18-1.67). **Figure 2A** illustrates -log10(*p*-values) of all metabolites derived from the univariable analysis, with the *p*-value of 0.0001 for myristoleic acid highlighted in red. **Figure 2B** presents a boxplot visually capturing the group-specific distribution of normalized values for myristoleic acid, along with the associated Wilcoxon *p*-value of 0.00055.

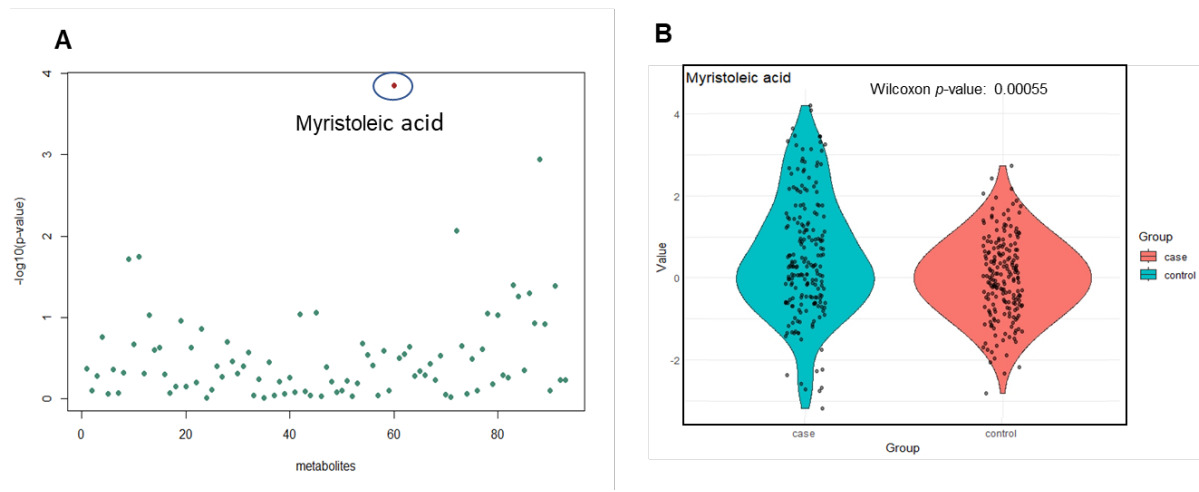

**Figure 2**: **A.** Representing the -log10(*p*-values) of all metabolites based on univariable analysis while adjusting for covariates. Myristoleic acid was found to be significant after

correction for multiple testing (FDR < 0.05). **B.** Representing the group-specific boxplot of myristoleic normalized values, with the Wilcoxon *p*-value of 0.00055.

**Discussion**

A metabolomic study of participants in The Epidemiological Research on Autism in Jamaica (ERAJ) study measured targeted metabolites in plasma using high-throughput liquid chromatography-mass spectrometry techniques, encompassing targeted metabolomic profiling including amino acids, amino sugar, fatty acids, central carbon metabolites, one carbon metabolites, nucleotides, and other key metabolic compounds. This case-control study identified four metabolites associated with ASD through simultaneous analysis of all 96 metabolites that passed the quality controls. Three of the findings involved fatty acids (myristoleic acid, octadecenoic acid, and eicosatetraenoic acid) and one amino acid was identified (sarcosine/alanine).

Myristoleic acid, classified as a monounsaturated omega-5 fatty acid, is not synthesized in sufficient quantities by the body and must be obtained through dietary sources. Potential anti-inflammatory properties associated with myristoleic acid have been reported[20]. However, we observed significantly elevated levels of myristoleic acid in a group of children diagnosed with ASD compared to TD controls. This elevation has been observed in a study involving Japanese children diagnosed with ASD[21]. This finding prompts further investigation into the metabolic pathways and potential implications of heightened myristoleic acid in the context of ASD and neurodevelopmental conditions.

Sarcosine/alanine is a nonproteinogenic amino acid that occurs as an intermediate product in the synthesis and degradation of the amino acid glycine. We observed an elevation of sarcosine associated with ASD, which has been previously reported in a study of Chinese children with ASD[22].

Eicosatetraenoic acid (ETA) belongs to the family of eicosanoids, signaling molecules derived from polyunsaturated fatty acids. These bioactive lipids play crucial roles in mediating cell-cell communication and may contribute to an anti-inflammatory response[23], consistent with its inverse association with ASD in our study.

Another fatty acid that we observed associated with ASD was octadecanoic acid which is essential for brain development. In a study investigating abnormalities of fatty acids and their impact on autism treatment, elevation of octadecanoic acid was linked to neurotoxicity in rats[24].

Increasing evidence supports the involvement of fatty acids in the etiology of ASD[3], and several studies have highlighted abnormalities in the lipid panel among ASD patients. The degree of these alterations potentially influences the severity of clinical symptoms[21,25]. In addition, a link between lipid metabolism and oxidative stress in ASD, which is closely related to inflammation, has been reported[21]. Intervention studies focusing on fatty acids, such as omega-3 supplementation, have yielded promising results, demonstrating improvements in symptoms such as irritability, hyperactivity, and social function in children with ASD[26–28]. Research indicates that children diagnosed with ASD tend to exhibit an elevated omega-6 to omega-3 ratio in their blood, which is associated with increased inflammation. This elevation might be due to reduced dietary intake or differences in fatty acid metabolism and cellular membrane incorporation, unique to ASD populations[27]. Alterations in membrane lipid components can influence crucial intra- and intercellular signaling pathways in various ways. For instance, lipids participate in the regulation of membrane-bound proteins involved in various neuronal processes, including synaptic transmission, signal transduction, and cell adhesion. These can contribute to aberrant neuronal signaling and synaptic dysfunction observed in individuals with ASD[21,29,30].

Given that the identified metabolites in this study fall into the category of non-essential compounds influenced by dietary habits, the observed differences in dietary patterns, as determined through the analysis of food consumption, warrant further investigation and replication. It is important to note that our analysis did not reveal any significant correlations between the identified metabolites and the information obtained from food questionnaires. This lack of correlation may be attributed to diet's intricate and multifaceted impact on metabolites. Additionally, the reliability of the scores derived from the questionnaires could introduce uncertainty into the analysis. Therefore, further exploration is required to fully comprehend the complex relationship between dietary habits and metabolite levels. In addition, given the heterogeneity among patients with ASD and variability within the ASD group, further investigation is warranted to understand individual-level differences. In future studies, investigating interactions among these metabolites could expand our understanding of the biological mechanisms underlying ASD[31,32]. Overall, these findings underscore the role of metabolites in the etiology of ASD and suggest potential avenues for the development of targeted interventions.

**Grant Numbers/Other Funding Source:**

National Institute of Environmental Health Sciences (NIEHS), # R01ES022165; Eunice Kennedy Shriver National Institute of Child Health and Human Development (NICHD) and National Institutes of Health Fogarty International Center (NICHD-FIC), # R21HD057808; National Center for Research Resources (NCRR), # UL1 RR024148; National Center for Advancing Translational Sciences (NCATS), # UL1 TR000371; National Center for Advancing Translational Sciences (NCATS), # UL1 TR003167. In addition, we acknowledge partial funding from the Department of Internal Medicine at the UTHealth Houston McGovern Medical School. The results reported here are those of the authors and don't necessary reflect the opinion of any of the institutions that provided the financial support.

**Acknowledgments**

The metabolomics core was supported by the CPRIT Core Facility Support Award RP210227 "Proteomic and Metabolomic Core Facility Facility (N.P.)", NIH (P30 CA125123), and intramural funds from the Dan L. Duncan Cancer Center (DLDCC).

**Authors' Conflict of Interest**

No conflict of interest

# Supplement

**Metabolomic profiles in Jamaican children with and without autism spectrum disorder**

**Figure S1. Batch effect correction.**

**A.** Data representation using first two principal components before the correction.

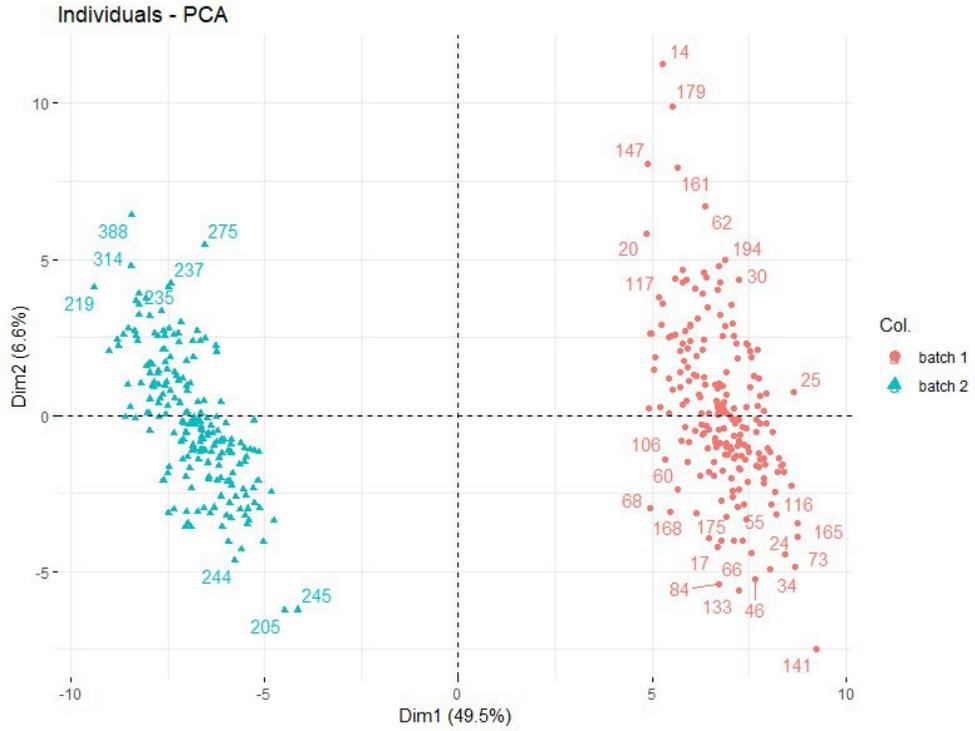

**B.** Data representation using first two principal components after the correction.

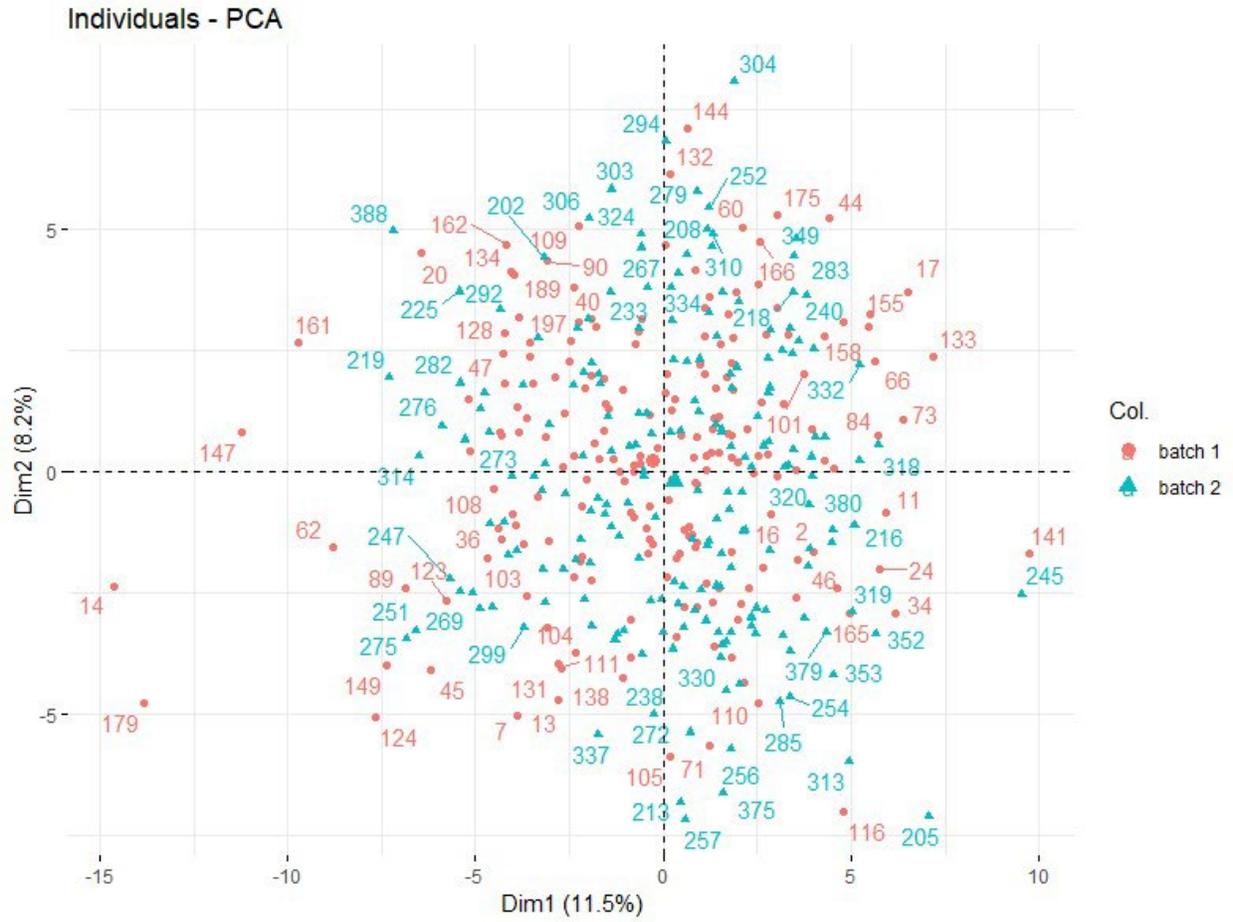

**Figure S2. Quality assessment.**

**A.** Metabolite clustering for the TD children group (controls).

**B.** Metabolite clustering for the ASD group (cases).

Table S1. By performing the Chi-square test or Fisher's exact test (*), we investigated the potential distinctions in dietary patterns between ASD and TD control groups based on food consumption scores obtained from questionnaires.

|  | TD control group (N=200) | ASD group (N=200) | Total (N=400) | *p*-value |
|---|---|---|---|---|
| **Sea fish** |  |  |  | 0.004 |
| Consumed | 133 (66.5%) | 105 (52.5%) | 238 (59.5%) |  |
| Not Consumed | 67 (33.5%) | 95 (47.5%) | 162 (40.5%) |  |
| **Fresh water** |  |  |  | 0.182 |
| Consumed | 49 (24.5%) | 38 (19.0%) | 87 (21.8%) |  |
| Not Consumed | 151 (75.5%) | 162 (81.0%) | 313 (78.2%) |  |
| **Sardine** |  |  |  | < 0.001 |
| Consumed | 170 (85.0%) | 139 (69.5%) | 309 (77.2%) |  |
| Not Consumed | 30 (15.0%) | 61 (30.5%) | 91 (22.8%) |  |
| **Tuna** |  |  |  | 0.196 |
| Consumed | 69 (34.5%) | 57 (28.5%) | 126 (31.5%) |  |
| Not Consumed | 131 (65.5%) | 143 (71.5%) | 274 (68.5%) |  |
| **Saltfish** |  |  |  | < 0.001 |
| Consumed | 158 (79.0%) | 113 (56.5%) | 271 (67.8%) |  |
| Not Consumed | 42 (21.0%) | 87 (43.5%) | 129 (32.2%) |  |
| **Shell fish lobster** |  |  |  | < 0.001 |
| Consumed | 33 (16.5%) | 6 (3.0%) | 39 (9.8%) |  |
| Not Consumed | 167 (83.5%) | 194 (97.0%) | 361 (90.2%) |  |
| **Shrimp** |  |  |  | < 0.001 |
| Consumed | 33 (16.5%) | 6 (3.0%) | 39 (9.8%) |  |
| Not Consumed | 167 (83.5%) | 194 (97.0%) | 361 (90.2%) |  |
| **Packaged fish** |  |  |  | 0.388 |
| Consumed | 38 (19.0%) | 45 (22.5%) | 83 (20.8%) |  |
| Not Consumed | 162 (81.0%) | 155 (77.5%) | 317 (79.2%) |  |
| **Beef** |  |  |  | 0.085 |
| Consumed | 92 (46.0%) | 75 (37.5%) | 167 (41.8%) |  |
| Not Consumed | 108 (54.0%) | 125 (62.5%) | 233 (58.2%) |  |
| **Lamb mutton** |  |  |  | 0.110 |
| Consumed | 33 (16.5%) | 22 (11.0%) | 55 (13.8%) |  |
| Not Consumed | 167 (83.5%) | 178 (89.0%) | 345 (86.2%) |  |
| **Goat** |  |  |  | < 0.001 |
| Consumed | 101 (50.5%) | 68 (34.0%) | 169 (42.2%) |  |

|  | TD control group (N=200) | ASD group (N=200) | Total (N=400) | p-value |
|---|---|---|---|---|
| Not Consumed | 99 (49.5%) | 132 (66.0%) | 231 (57.8%) | |
| **Pork** | | | | < 0.001 |
| Consumed | 112 (56.0%) | 79 (39.5%) | 191 (47.8%) | |
| Not Consumed | 88 (44.0%) | 121 (60.5%) | 209 (52.2%) | |
| **Liver** | | | | 0.009 |
| Consumed | 117 (58.5%) | 91 (45.5%) | 208 (52.0%) | |
| Not Consumed | 83 (41.5%) | 109 (54.5%) | 192 (48.0%) | |
| **Chicken** | | | | 0.006* |
| Consumed | 196 (98.0%) | 184 (92.0%) | 380 (95.0%) | |
| Not Consumed | 4 (2.0%) | 16 (8.0%) | 20 (5.0%) | |
| **Milk** | | | | 0.229 |
| Consumed | 98 (49.0%) | 86 (43.0%) | 184 (46.0%) | |
| Not Consumed | 102 (51.0%) | 114 (57.0%) | 216 (54.0%) | |
| **Cheese** | | | | < 0.001 |
| Consumed | 165 (82.5%) | 134 (67.0%) | 299 (74.8%) | |
| Not Consumed | 35 (17.5%) | 66 (33.0%) | 101 (25.2%) | |
| **Yogurt** | | | | 0.004 |
| Consumed | 68 (34.0%) | 42 (21.0%) | 110 (27.5%) | |
| Not Consumed | 132 (66.0%) | 158 (79.0%) | 290 (72.5%) | |
| **Eggs** | | | | < 0.001 |
| Consumed | 172 (86.0%) | 144 (72.0%) | 316 (79.0%) | |
| Not Consumed | 28 (14.0%) | 56 (28.0%) | 84 (21.0%) | |
| **Rice** | | | | < 0.001* |
| Consumed | 199 (99.5%) | 178 (89.0%) | 377 (94.2%) | |
| Not Consumed | 1 (0.5%) | 22 (11.0%) | 23 (5.8%) | |
| **Fried dumpling** | | | | 0.140 |
| Consumed | 154 (77.0%) | 141 (70.5%) | 295 (73.8%) | |
| Not Consumed | 46 (23.0%) | 59 (29.5%) | 105 (26.2%) | |
| **Boiled dumpling** | | | | < 0.001 |
| Consumed | 184 (92.0%) | 154 (77.0%) | 338 (84.5%) | |
| Not Consumed | 16 (8.0%) | 46 (23.0%) | 62 (15.5%) | |
| **White bread** | | | | < 0.001 |
| Consumed | 140 (70.0%) | 103 (51.5%) | 243 (60.8%) | |
| Not Consumed | 60 (30.0%) | 97 (48.5%) | 157 (39.2%) | |
| **Whole wheat bread** | | | | 0.760 |
| Consumed | 121 (60.5%) | 118 (59.0%) | 239 (59.8%) | |
| Not Consumed | 79 (39.5%) | 82 (41.0%) | 161 (40.2%) | |

|  | TD control group (N=200) | ASD group (N=200) | Total (N=400) | p-value |
|---|---|---|---|---|
| **Cakes bun** |  |  |  | 0.003 |
| Consumed | 164 (82.0%) | 138 (69.0%) | 302 (75.5%) |  |
| Not Consumed | 36 (18.0%) | 62 (31.0%) | 98 (24.5%) |  |
| **Porridge** |  |  |  | 0.003 |
| Consumed | 184 (92.0%) | 164 (82.0%) | 348 (87.0%) |  |
| Not Consumed | 16 (8.0%) | 36 (18.0%) | 52 (13.0%) |  |
| **Cold cereal** |  |  |  | < 0.001 |
| Consumed | 157 (78.5%) | 116 (58.0%) | 273 (68.2%) |  |
| Not Consumed | 43 (21.5%) | 84 (42.0%) | 127 (31.8%) |  |
| **Macaroni** |  |  |  | < 0.001 |
| Consumed | 179 (89.5%) | 150 (75.0%) | 329 (82.2%) |  |
| Not Consumed | 21 (10.5%) | 50 (25.0%) | 71 (17.8%) |  |
| **Peas** |  |  |  | < 0.001 |
| Consumed | 146 (73.0%) | 106 (53.0%) | 252 (63.0%) |  |
| Not Consumed | 54 (27.0%) | 94 (47.0%) | 148 (37.0%) |  |
| **Beans** |  |  |  | 0.011 |
| Consumed | 94 (47.0%) | 69 (34.5%) | 163 (40.8%) |  |
| Not Consumed | 106 (53.0%) | 131 (65.5%) | 237 (59.2%) |  |
| **Nuts** |  |  |  | < 0.001 |
| Consumed | 158 (79.0%) | 95 (47.5%) | 253 (63.2%) |  |
| Not Consumed | 42 (21.0%) | 105 (52.5%) | 147 (36.8%) |  |
| Yam |  |  |  | 0.008 |
| Consumed | 147 (73.5%) | 122 (61.0%) | 269 (67.2%) |  |
| Not Consumed | 53 (26.5%) | 78 (39.0%) | 131 (32.8%) |  |
| **Carrot** |  |  |  | < 0.001 |
| Consumed | 174 (87.0%) | 147 (73.5%) | 321 (80.2%) |  |
| Not Consumed | 26 (13.0%) | 53 (26.5%) | 79 (19.8%) |  |
| **Lettuce** |  |  |  | < 0.001 |
| Consumed | 107 (53.5%) | 57 (28.5%) | 164 (41.0%) |  |
| Not Consumed | 93 (46.5%) | 143 (71.5%) | 236 (59.0%) |  |
| **Callaloo** |  |  |  | 0.001 |
| Consumed | 153 (76.5%) | 123 (61.5%) | 276 (69.0%) |  |
| Not Consumed | 47 (23.5%) | 77 (38.5%) | 124 (31.0%) |  |
| **Cabbage** |  |  |  | < 0.001 |
| Consumed | 148 (74.0%) | 109 (54.5%) | 257 (64.2%) |  |
| Not Consumed | 52 (26.0%) | 91 (45.5%) | 143 (35.8%) |  |
| **String beans** |  |  |  | 0.020 |

|  | TD control group (N=200) | ASD group (N=200) | Total (N=400) | *p*-value |
|---|---|---|---|---|
| Consumed | 59 (29.5%) | 39 (19.5%) | 98 (24.5%) |  |
| Not Consumed | 141 (70.5%) | 161 (80.5%) | 302 (75.5%) |  |
| **Tomatoes** |  |  |  | < 0.001 |
| Consumed | 142 (71.0%) | 105 (52.5%) | 247 (61.8%) |  |
| Not Consumed | 58 (29.0%) | 95 (47.5%) | 153 (38.2%) |  |
| **Ackee** |  |  |  | < 0.001 |
| Consumed | 155 (77.5%) | 94 (47.0%) | 249 (62.2%) |  |
| Not Consumed | 45 (22.5%) | 106 (53.0%) | 151 (37.8%) |  |
| **Avocado** |  |  |  | < 0.001 |
| Consumed | 106 (53.0%) | 42 (21.0%) | 148 (37.0%) |  |
| Not Consumed | 94 (47.0%) | 158 (79.0%) | 252 (63.0%) |  |
| **Green banana** |  |  |  | < 0.001 |
| Consumed | 155 (77.5%) | 123 (61.5%) | 278 (69.5%) |  |
| Not Consumed | 45 (22.5%) | 77 (38.5%) | 122 (30.5%) |  |
| **Fried plantain** |  |  |  | < 0.001 |
| Consumed | 173 (86.5%) | 142 (71.0%) | 315 (78.8%) |  |
| Not Consumed | 27 (13.5%) | 58 (29.0%) | 85 (21.2%) |  |
| **Ripe banana** |  |  |  | < 0.001 |
| Consumed | 189 (94.5%) | 162 (81.0%) | 351 (87.8%) |  |
| Not Consumed | 11 (5.5%) | 38 (19.0%) | 49 (12.2%) |  |
| **Oranges** |  |  |  | < 0.001 |
| Consumed | 181 (90.5%) | 134 (67.0%) | 315 (78.8%) |  |
| Not Consumed | 19 (9.5%) | 66 (33.0%) | 85 (21.2%) |  |
| **Tangerine** |  |  |  | < 0.001 |
| Consumed | 95 (47.5%) | 52 (26.0%) | 147 (36.8%) |  |
| Not Consumed | 105 (52.5%) | 148 (74.0%) | 253 (63.2%) |  |
| **Grapes** |  |  |  | < 0.001 |
| Consumed | 123 (61.5%) | 79 (39.5%) | 202 (50.5%) |  |
| Not Consumed | 77 (38.5%) | 121 (60.5%) | 198 (49.5%) |  |
| **Otaheite apples** |  |  |  | < 0.001 |
| Consumed | 169 (84.5%) | 128 (64.0%) | 297 (74.2%) |  |
| Not Consumed | 31 (15.5%) | 72 (36.0%) | 103 (25.8%) |  |
| **Pineapples** |  |  |  | < 0.001 |
| Consumed | 143 (71.5%) | 90 (45.0%) | 233 (58.2%) |  |
| Not Consumed | 57 (28.5%) | 110 (55.0%) | 167 (41.8%) |  |
| **American apples** |  |  |  | 0.002 |
| Consumed | 78 (39.0%) | 49 (24.5%) | 127 (31.8%) |  |

|  | TD control group (N=200) | ASD group (N=200) | Total (N=400) | p-value |
|---|---|---|---|---|
| Not Consumed | 122 (61.0%) | 151 (75.5%) | 273 (68.2%) |  |
| **Guinep** |  |  |  | 0.041 |
| Consumed | 12 (6.0%) | 4 (2.0%) | 16 (4.0%) |  |
| Not Consumed | 188 (94.0%) | 196 (98.0%) | 384 (96.0%) |  |
| **Peach** |  |  |  | 0.041* |
| Consumed | 12 (6.0%) | 4 (2.0%) | 16 (4.0%) |  |
| Not Consumed | 188 (94.0%) | 196 (98.0%) | 384 (96.0%) |  |
| **Plums** |  |  |  | < 0.001 |
| Consumed | 98 (49.0%) | 24 (12.0%) | 122 (30.5%) |  |
| Not Consumed | 102 (51.0%) | 176 (88.0%) | 278 (69.5%) |  |
| **Strawberry** |  |  |  | 0.388 |
| Consumed | 21 (10.5%) | 16 (8.0%) | 37 (9.2%) |  |
| Not Consumed | 179 (89.5%) | 184 (92.0%) | 363 (90.8%) |  |
| **Naseberry** |  |  |  | 0.004 |
| Consumed | 70 (35.0%) | 44 (22.0%) | 114 (28.5%) |  |
| Not Consumed | 130 (65.0%) | 156 (78.0%) | 286 (71.5%) |  |
| **Sweetsop** |  |  |  | < 0.001 |
| Consumed | 69 (34.5%) | 23 (11.5%) | 92 (23.0%) |  |
| Not Consumed | 131 (65.5%) | 177 (88.5%) | 308 (77.0%) |  |
| **Mango** |  |  |  | < 0.001 |
| Consumed | 186 (93.0%) | 148 (74.0%) | 334 (83.5%) |  |
| Not Consumed | 14 (7.0%) | 52 (26.0%) | 66 (16.5%) |  |
| **June plum** |  |  |  | < 0.001 |
| Consumed | 142 (71.0%) | 74 (37.0%) | 216 (54.0%) |  |
| Not Consumed | 58 (29.0%) | 126 (63.0%) | 184 (46.0%) |  |
| **Juices** |  |  |  | < 0.001 |
| Consumed | 186 (93.0%) | 161 (80.5%) | 347 (86.8%) |  |
| Not Consumed | 14 (7.0%) | 39 (19.5%) | 53 (13.2%) |  |
| **Beverages** |  |  |  | 0.514 |
| Consumed | 181 (90.5%) | 177 (88.5%) | 358 (89.5%) |  |
| Not Consumed | 19 (9.5%) | 23 (11.5%) | 42 (10.5%) |  |
| **Soft drinks** |  |  |  | 0.057 |
| Consumed | 104 (52.0%) | 85 (42.5%) | 189 (47.2%) |  |
| Not Consumed | 96 (48.0%) | 115 (57.5%) | 211 (52.8%) |  |
| **Tea substitutes** |  |  |  | < 0.001 |
| Consumed | 185 (92.5%) | 160 (80.0%) | 345 (86.2%) |  |
| Not Consumed | 15 (7.5%) | 40 (20.0%) | 55 (13.8%) |  |

|  | TD control group (N=200) | ASD group (N=200) | Total (N=400) | p-value |
|---|---|---|---|---|
| **Canned food** |  |  |  | < 0.001 |
| Consumed | 162 (81.0%) | 122 (61.0%) | 284 (71.0%) |  |
| Not Consumed | 38 (19.0%) | 78 (39.0%) | 116 (29.0%) |  |
| **Aluminum foil** |  |  |  | 0.303 |
| Consumed | 41 (20.5%) | 33 (16.5%) | 74 (18.5%) |  |
| Not Consumed | 159 (79.5%) | 167 (83.5%) | 326 (81.5%) |  |
| **Unpeeled fruits** |  |  |  | < 0.001 |
| Consumed | 179 (89.5%) | 124 (62.0%) | 303 (75.8%) |  |
| Not Consumed | 21 (10.5%) | 76 (38.0%) | 97 (24.2%) |  |
| **Animal fat** |  |  |  | 0.003 |
| Consumed | 44 (22.0%) | 22 (11.0%) | 66 (16.5%) |  |
| Not Consumed | 156 (78.0%) | 178 (89.0%) | 334 (83.5%) | 0.004 |